\documentclass[aps,prb,article,twocolumn,amsmath,amssymb,superscriptaddress]{revtex4-2}
\pdfmapline{-dummy LMRoman12-Regular}
\date{\today}
\usepackage{epsfig}
\usepackage{subfigure}
\usepackage{graphicx}
\usepackage{dcolumn}
\usepackage{bm}
\usepackage[colorlinks,linkcolor=blue,hyperindex,CJKbookmarks]{hyperref}
\usepackage{float}
\usepackage{hyperref}
\usepackage{comment}
\usepackage[utf8]{inputenc}

\usepackage[dvipsnames]{xcolor}

\begin{document}

\title{Superconducting Phases of the Square-Lattice Extended Hubbard Model}

\author{Wei-Chih Chen}
\affiliation{Department of Physics, University of Alabama at Birmingham, Birmingham, Alabama 35294, USA}
\affiliation{Department of Physics and Astronomy, Clemson University, Clemson, South Carolina 29631, USA}

\author{Yao Wang}
\email{yao.wang@emory.edu}
\affiliation{Department of Physics and Astronomy, Clemson University, Clemson, South Carolina 29631, USA}
\affiliation{Department of Chemistry, Emory University, Atlanta, GA, 30322, United States}

\author{Cheng-Chien Chen}
\email{chencc@uab.edu}
\affiliation{Department of Physics, University of Alabama at Birmingham, Birmingham, Alabama 35294, USA}

\begin{abstract}
We study the square-lattice extended Hubbard model with on-site $U$ and nearest-neighbor $V$ interactions by exact diagonalization. We show that non-equilibrium quench dynamics can help determine the equilibrium phase transition boundaries, which agree with the calculations of fidelity metric, dynamical structure factor, and correlation function. At half filling, the phase diagrams in the strong-coupling regime include spin density wave and $d_{x^2-y^2}$-wave superconductivity at large positive $U$, charge density wave (extended $s^*$-wave superconductivity) at large positive (negative) $V$, and $s$-wave superconductivity at large negative $U$ with vanishing $V$. The energies of different particle sectors also help determine the phase separation region. With carrier doping, charge fluctuation result in strong competition between different orders, making it more difficult to identify the leading instability on finite-size cluster. Nevertheless, the more exotic $p$-wave superconducting pairing is found to be enhanced when the system is heavily overdoped by $37.5\%-50\%$ holes, especially in interaction parameter range relevant to the cuprate superconductors.
\end{abstract}

\date{\today}
\maketitle

\section{Introduction}
One of the most intriguing but challenging topics in condensed matter concerns studying correlation effects in many interacting particles. Strongly correlated systems exhibit various symmetry-breaking states, such as superconductivity, which can arise from the interplay of spin, charge, and lattice degrees of freedom\,\cite{dagotto1994correlated, davis2013concepts}. Already in 1950s, the Bardeen–Cooper–Schrieffer theory successfully explained the mechanism of conventional superconductors through phonon-mediated electron pairs\,\cite{bardeen1957microscopic}. In late 1980s, the discovery of copper-based high-temperature superconductors -- later identified with a $d$-wave pairing symmetry\,\cite{tsuei2000pairing} -- reformed our understanding of correlated materials\,\cite{bednorz1986possible}. Although the ultimate theory for high-temperature superconductivity remains controversial, the strong electron repulsive interactions in copper $3d$ orbitals are believed to play crucial roles\,\cite{anderson1987resonating,zhang1988effective,lee2006doping}.
Unconventional superconductivity also has been explored in other transition-metal oxides. For example, the infinite-layer nickelates share similar electronic structures with the cuprates and become superconducting at $\sim$20\% hole doping\,\cite{li2019superconductivity}. Moreover, superconductivity in the ruthenates was identified with a triplet $p$-wave pairing\,\cite{maeno1994superconductivity,rice1995sr2ruo4,baskaran1996sr2ruo4,mackenzie2003superconductivity,nelson2004odd}, although the claim was challenged by recent experiments\,\cite{pustogow2019constraints,ishida2020reduction}. In any case, these unconventional superconducting phases are believed to originate from strong electron correlation effects.

The simplest toy model to describe electronic correlation is the single-band Hubbard model with a local on-site interaction $U$. This simple model already can explain the behaviors of Mott insulator, stripe order\,\cite{zheng2017stripe, huang2017numerical,huang2018stripe,ponsioen2019period}, strange metal\,\cite{kokalj2017bad, huang2019strange, cha2020slope}, and to some extent $d$-wave superconductivity\,\cite{maier2005systematic, zheng2016ground,ido2018competition,jiang2019superconductivity}.
As a natural extension, more recent studies have considered a non-local Coulomb interaction (denoted as $V$) in the so-called extended Hubbard model (EHM).
A repulsive $V$ can induce a charge density wave (CDW), and an attractive $V$ was argued to favor spin-triplet superconductivity\,\cite{lin1986condensation,penc1994phase}. Already in 1D systems, a $p$-wave superconducting phase was predicted for repulsive $U$ and attractive $V$ in EHM\,\cite{lin1995phase,lin1997phase,xiang2019doping, shinjo2019machine, qu2022spin}. Similar studies have been explored in 2D systems\,\cite{wahle1998microscopic, onari2004phase,aichhorn2004charge, huang2013unconventional, nayak2018exotic,van2018extended}, using weak-coupling theory or with approximation in treating interaction effect. To understand and design exotic phases like $p$-wave superconductivity, it is necessary to study EHM in the strong-coupling regime.

Moreover, recent ARPES experiments on 1D cuprate chains have revealed a sizeable attractive Coulomb interaction between nearest-neighbor (NN) electrons\,\cite{chen2021anomalously}. 
Although not as strong as the on-site Coulomb repulsion, this attractive interaction is comparable to the electron hopping integral, and thereby should not be ignored\,\cite{li2022suppressed}.
The structural similarities among the cuprates also suggest that attractive NN interaction may exist in quasi-2D cuprate materials in general.
Since electron-phonon coupling may be its origin, this attractive interaction should be even stronger in a 2D copper-oxide plane, because of the richer phonon modes and stronger impact of ligands. Therefore, studying the 2D EHM could be important for understanding high-$T_c$ pairing mechanism.

\begin{figure}[!t]
\begin{center}
\includegraphics[width=1\columnwidth]{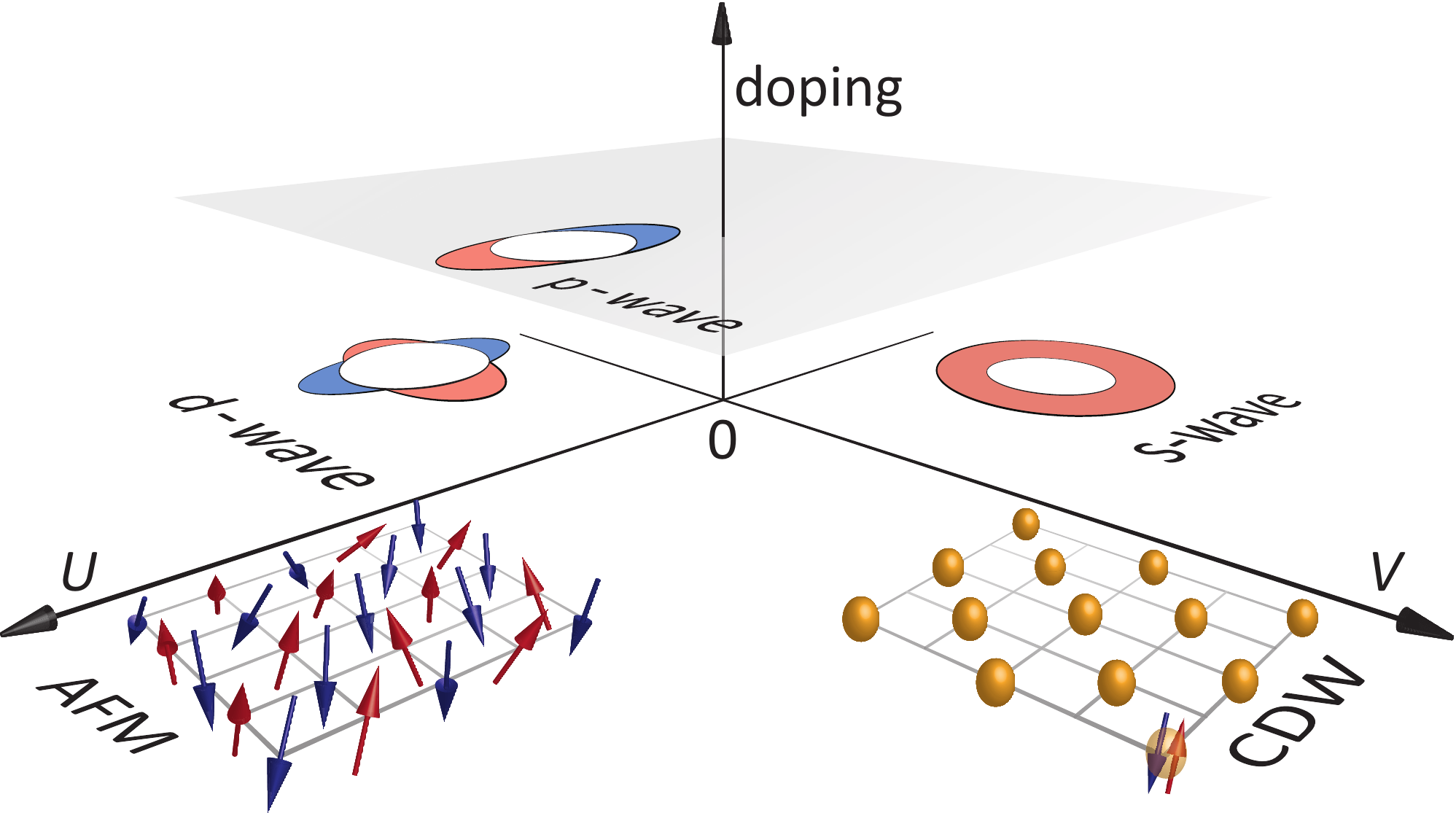}
\caption{\label{fig:cartoon}
Schematic phase diagram of the square-lattice extended Hubbard model. Depending on the onsite $U$ and nearest-neighbor $V$ interactions, as well as hole doping levels, the model can support various symmetry breaking phases, such as spin density wave (SDW), charge density wave (CDW), $s$-wave superconductivity, $d_{x^2-y^2}$-wave superconductivity, and the more exotic $p$-wave superconductivity. A phase separation (PS) also can occur near the $d$-wave phase especially with negative $V$.
}
\end{center}
\end{figure}

In this paper, we systematically study the phase diagrams of the square-lattice EHM as functions of interaction strengths and doping levels.
Studying strong correlation effect is challenging, and most theoretical and computational methods are based on various levels of approximation in handling the interaction.
Here we choose to use exact diagonalization (ED), which has the advantage that electron interaction effect can be treated exactly.
We employ ED to compute quench dynamics, fidelity metrics, and various correlation functions.
While the ED study is restricted to finite-size clusters, we will show that the phase boundaries determined by ED agree with the functional renormalization group (fRG) results in the weak-coupling limit\,\cite{huang2013unconventional},
and our approach remains valid in the strong-coupling limit.
Figure \ref{fig:cartoon} illustrates the main results of our study.
At half filling, spin density wave (SDW) is dominant at positive $U$, and $s$-wave superconductivity is dominant at negative $U$ and small $|V|$. On the other hand, a positive $V$ induces a charge density wave (CDW), and a negative $V$ can cause phase separation. When the system is heavily overdoped by hole carriers, the more exotic $p$-wave superconducting correlation can be enhanced at large positive $U$ and small negative $V$. We also determine the phase  separated  region by computing the energies of different particle sectors. These results will be relevant to various physical systems such as the cuprates and cold-atoms experiments\,\cite{mazurenko2017cold, bohrdt2021exploration}.

\section{Model and Methods}
The extended Hubbard model (EHM) reads
\begin{eqnarray}\label{eqHam}
H = -t_h \sum_{\langle ij \rangle, \sigma} (c^\dagger_{i\sigma} c_{j\sigma} + h.c.)
+ U \sum_i n_{i\uparrow} n_{i\downarrow} + V\sum_{\substack{\langle ij \rangle,\\ \sigma\sigma'}} n_{i\sigma} n_{j\sigma'}.\nonumber\\
\end{eqnarray}
Here, $c^\dagger_{i\sigma}$ creates an electron with spin $\sigma$ (= $\uparrow$ or $\downarrow$) at lattice site $i$, and $n_{i\sigma} = c^\dagger_{i\sigma} c_{i\sigma}$ is the corresponding fermion number operator. $\langle ... \rangle$ represents a pair of near-neighbor (NN) sites. $t_h$ is the hopping amplitude. $U$ and $V$ are respectively the on-site and NN interaction strengths; a positive (negative) value corresponds to repulsive (attractive) force. Here, $t_h \equiv 1$ and we explore both positive and negative interactions in the range $|U| \le 10$ and $|V| \le 10$. We note that a small, attractive NN Coulomb interaction $V$ has been identified in ARPES experiments in doped cuprate chains\,\cite{chen2021anomalously}. While longer-range interactions beyond NN might also be present, their strengths are expected to be weaker and remain to be established experimentally. We solve Eq.~\eqref{eqHam} using the ED technique, which can treat correlation effects exactly for arbitrary interaction strength. The main challenge of ED is that the Hilbert space size grows exponentially with the lattice size, so the calculations are limited to finite-size clusters. Here, we focus on a $4\times 4$ square cluster (under periodic boundary condition), which exhibits the proper $D_4$ symmetry for studying the competition between $d-$wave and $p$-wave superconductivity. On this $N=16$ site cluster, we study hole doping levels at 0\% (half filling with 1 electron per site on average), 12.5\%, 25\%, 37.5\%, and 50\% (quarter filling with 0.5 electron per site on average).

The ED algorithm proceeds as follows. We first construct the Hamiltonian matrix using an eigenbasis of the fermion number operator. We then perform matrix diagonalization to obtain the ground state $|G_{\lambda}\rangle$ with interaction parameters $\lambda \equiv (U, V)$. Depnding on $\lambda$ and the doping level, ground state degeneracy can occur, and these degenerate states have to be considered when computing the expectation values of physical observables. The actual diagonalization is performed using iterative Krylov sub-space methods as implemented in the PETSc\,\cite{petsc-web-page, petsc-efficient} and SLEPc\,\cite{Hernandez:2003:SSL,Hernandez:2005:SSF} libraries. The Krylov-Schur technique is utilized to resolve degenerate eigenstates.

After obtaining the ground state(s) $|G_{\lambda}\rangle$ for a Hamiltonian $H_0$ with interactions $\lambda = (U, V)$, we next perform an interaction quench, where the Hamiltonian undergoes a sudden change to $H_1$ with interactions $\lambda^* = (U^*,V^*)$ at time $t = 0^+$. The state at time $t+\delta t$ is obtained by acting the time evolution operator on the state at time $t$:
\begin{equation}
    |\psi (t + \delta t)\rangle = e^{-iH_1 \delta t} |\psi(t)\rangle, \,\textrm{for } t\ge 0.
\end{equation}
Here, $|\psi (t=0) \rangle \equiv |G_\lambda\rangle$ is the ground state of $H_0$. The time evolution calculation is again performed using a Krylov method in PETSc, which requires only repeated matrix-vector multiplications to construct the Krylov subspace $\{ | \psi (t)\rangle, H^1_1| \psi (t)\rangle, H^2_1| \psi (t)\rangle, H^3_1| \psi (t)\rangle, ... \}$, {\it without} the need to explicitly construct the matrix exponential operator $e^{-iH_1 \delta t}$. In our calculations, we typically evolve the system to a final time $t =  50  {t_h}^{-1}$, with a time step $\delta t = 0.01 - 0.05 {t_h}^{-1}$. For selected sets of interaction parameters, we have performed additional calculations with varying time steps to ensure that the numerical error due to a finite time discretization is negligible.

With $|\psi (t) \rangle$ available, we compute the equal-time correlation functions $\langle \psi (t) | \hat{O}^\dagger \hat{O} | \psi(t)\rangle$ for different order parameters. In particular, the correlations with the following operators $\hat O$ are predominant at half filling:
\begin{eqnarray}
\rho_\mathbf{q} &=& \frac{1}{\sqrt{N}}\sum_{i} e^{i\mathbf{q}\cdot\mathbf{r}_i} (c^{\dagger}_{i\uparrow}c_{i\uparrow} + c^{\dagger}_{i\downarrow}c_{i\downarrow}),\\
\rho^s_\mathbf{q} &=& \frac{1}{2\sqrt{N}}\sum_{i} e^{i\mathbf{q}\cdot\mathbf{r}_i} (c^{\dagger}_{i\uparrow}c_{i\uparrow} - c^{\dagger}_{i\downarrow}c_{i\downarrow} ),\\
\Delta_s &=& \frac{1}{\sqrt{N}}\sum_i c_{i\uparrow}c_{i\downarrow},\\
\Delta_{d_{x^2-y^2}} &=& \frac{1}{2\sqrt{N}} \sum_i (c_{i\uparrow}c_{i+\hat{x} \downarrow} + c_{i\uparrow}c_{i-\hat{x} \downarrow}  \\\nonumber 
&-& c_{i\uparrow}c_{i+\hat{y} \downarrow} - c_{i\uparrow}c_{i-\hat{y} \downarrow}).
\end{eqnarray}
Here, $\rho_{\mathbf{q}}$ ($\rho_{\mathbf{q}}^{s}$) is the charge (spin) density operator in momentum space, relevant to a CDW (SDW) phase at ordering vector $\mathbf{q}$. $\Delta_s$ and $\Delta_{d_{x^2-y^2}}$ are real-space $s$-wave and $d_{x^2-y^2}$-wave pairing operators, respectively. Upon doping, other superconducting instabilities can occur, and we also study the pairing operators for extended $s$-wave ($s^*$-wave), $d_{xy}$-wave, and $p_x$-wave superconductivity:
\begin{eqnarray}
\Delta_{s^*} &=& \frac{1}{2\sqrt{N}} \sum_i ( c_{i\uparrow}c_{i+\hat{x} \downarrow} + c_{i\uparrow}c_{i-\hat{x} \downarrow} \\\nonumber  
&+& c_{i\uparrow}c_{i+\hat{y} \downarrow} + c_{i\uparrow}c_{i-\hat{y} \downarrow}),\\
\Delta_{d_{xy}}&=& \frac{1}{2\sqrt{N}} \sum_i (c_{i\uparrow}c_{i+\hat{x} +\hat{y} \downarrow} + c_{i\uparrow}c_{i-\hat{x} -\hat{y} \downarrow} \\\nonumber &-& c_{i\uparrow}c_{i-\hat{x}+\hat{y} \downarrow} - c_{i\uparrow}c_{i+\hat{x}-\hat{y} \downarrow}),\\
\Delta_{p_x} &=& \frac{1}{\sqrt{2 N}} \sum_i (c_{i\uparrow}c_{i+\hat{x} \downarrow} - c_{i\uparrow}c_{i-\hat{x} \downarrow}).
\end{eqnarray}
By symmetry, the value of $p_y$-wave correlation is identical to that of $p_x$-wave in our ED calculations. After obtaining $\langle \psi (t) | \hat{O}^\dagger \hat{O} | \psi(t)\rangle$ for the above order parameters, we Fourier transform the time-domain data to extract the spectral features in frequency space. As shown later, the Fourier spectra are closely related to the equilibrium system's charge and spin excitation energies, which exhibit distinct behaviors in different broken-symmetry phases.

To extract the charge and spin gaps of the equilibrium system, we also compute the charge $N(\mathbf{q},\omega)$ and spin $S(\mathbf{q},\omega)$ dynamical structure factors:
\begin{eqnarray}
	N(\mathbf{q},\omega)\! &=&\! \frac1{\pi} \mathrm{Im}\left\langle G_{\lambda*}\! \left| \rho_{-\mathbf{q}} \frac1{H_1\! -\! E_{G^*} \!-\! \omega\! -\! i\Gamma}\rho_\mathbf{q} \right|\!G_{\lambda*}\!\right\rangle,\\
	S(\mathbf{q},\omega) &=& \frac1{\pi} \mathrm{Im}\left\langle \!G_{\lambda*}\! \left| \rho^s_{-\mathbf{q}} \frac1{H_1\! - \!E_{G^*} \!- \!\omega \!- \!i\Gamma}\rho^s_\mathbf{q} \right|\!G_{\lambda*}\!\right\rangle.
\end{eqnarray}
Here, $|G_{\lambda^*}\rangle$ is the equilibrium ground state of $H_1$ with energy $E_{G^*}$, and $\Gamma$ is a finite spectral broadening chosen to be $0.1t_h$ in this study.

Finally, we also compute the fidelity of quantum state overlap $f(\lambda^*,\lambda)\equiv|\langle G_{\lambda^*} | G_{\lambda}\rangle|$, where $|G_{\lambda}\rangle$ is the ground state with interactions $\lambda \equiv (U,V)$, and $\lambda^* \equiv \lambda + \delta \lambda$ represents a small deviation from $\lambda$. If ground state degeneracy occurs, the fidelity value can be obtained by a singular value decomposition of the overlap matrix with matrix elements $M_{k\ell}\equiv \langle G_{\lambda^*, k} | G_{\lambda,\ell} \rangle$. Here, $|G_{\lambda, \ell}\rangle$ ($|G_{\lambda^*, k}\rangle$) is the $\ell$-th ($k$-th) degenerate ground state for interaction $\lambda$ ($\lambda^*$), and the largest singular value is chosen to represent the fidelity.
The fidelity approach is powerful for determining phase transitions in quantum many-body systems\,\cite{gu2010fidelity,jia2011fidelity}. In particular, if $|G_{\lambda}\rangle$ and $|G_{\lambda^*}\rangle$ belong to the same broken-symmetry phase, $f$ is expected to be equal or close to unity. On the other hand, if $|G_{\lambda}\rangle$ and $|G_{\lambda^*}\rangle$ have a distinct nature, $f$ will be reduced from unity, which thereby signals a quantum phase transition with a small change $\delta \lambda$ in the interaction parameters. Using quench dynamics, spectral gaps, and fidelity metrics, we are able to probe quantum phase transition boundaries using ED with results in agreement with fRG in the weak-coupling limit, and to propose phase diagrams in the strong-coupling regime.

\section{Results and Discussion}

\begin{figure*}[!th]
\begin{center}
\includegraphics[width=16cm]{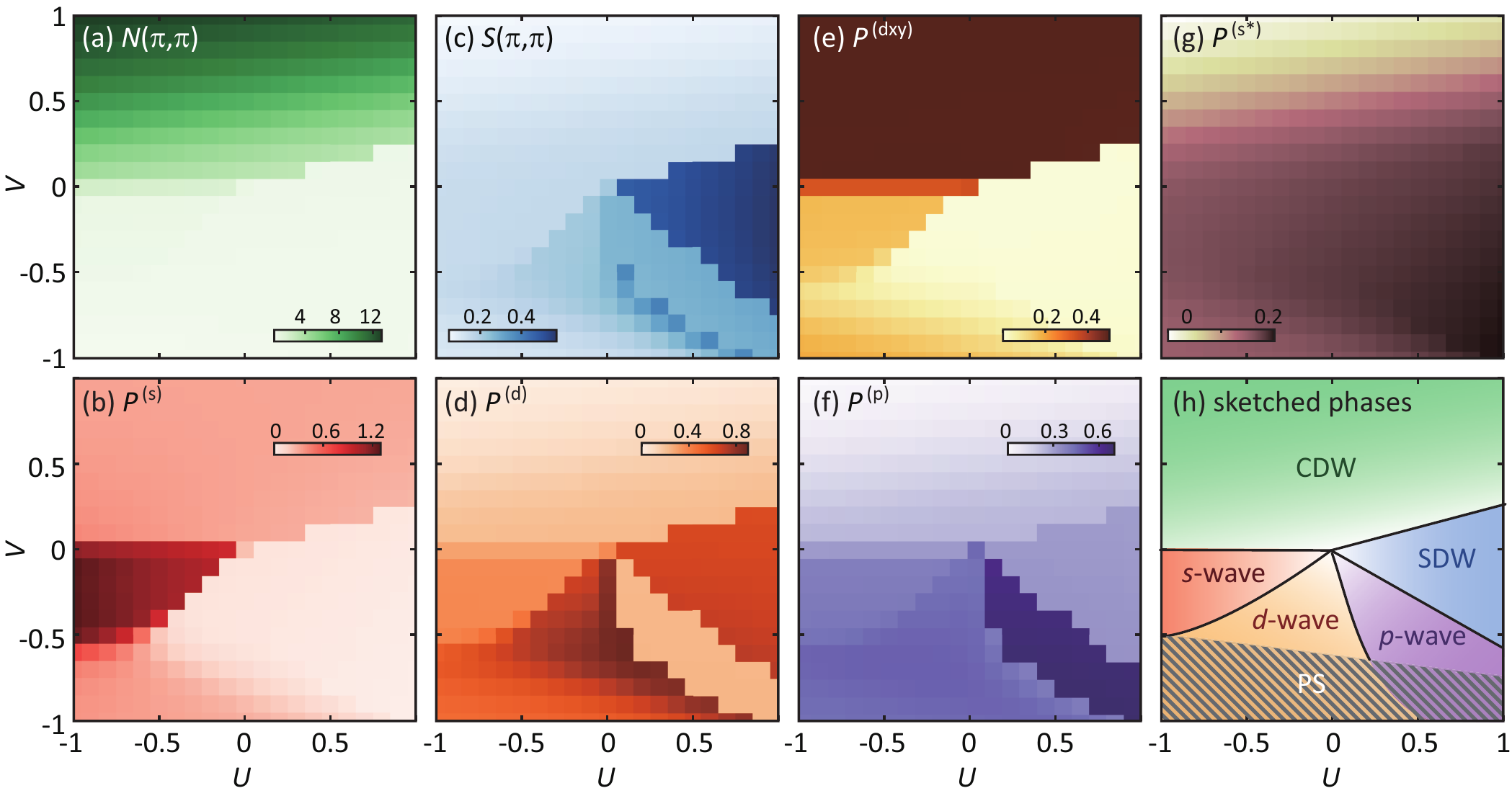}\vspace{-3mm}
\caption{\label{fig:correlationsWeakCoupling}
Correlation functions for different order parameters computed by exact diagonalization (ED) for half-filled extended Hubbard model in the weak-coupling regime: $|U| \le 1$ and $|V| \le 1$. Panel (h) depicts the corresponding phase diagram with phase boundaries determined by ED calculations of fidelity metric, correlation function, and total energy.
Gray dashed lines indicate regions where phase separation (PS) would occur.
}
\end{center}
\end{figure*}

\subsection{Weak-Coupling Limit}

We first focus on a small interaction range $|U| \le 1$ and $|V| \le 1$, where we can benchmark ED against fRG\,\cite{huang2013unconventional, metzner2012functional}. Figure \ref{fig:correlationsWeakCoupling} shows equilibrium correlation functions $\langle G_\lambda | \hat O^\dagger \hat O|G_\lambda\rangle$  for different order parameters in Eqs. (3)-(9) as functions of interaction strengths $\lambda = (U, V)$. The phase separation (PS) region is determined from the energy difference between different particle sectors (see
\hyperref[appendix:PS]{Appendix}).
Overall, a positive $V > 0$ will stabilize a $(\pi,\pi)$ charge density wave (CDW), as manifested in Fig.~\ref{fig:correlationsWeakCoupling}(a). Moreover, $V > 0$ with $U < 0$ will further enhance the tendency towards the CDW state. In contrast, a predominant positive $U > 0$ is expected to support a $(\pi,\pi)$ spin density wave (SDW), as seen in Fig.~\ref{fig:correlationsWeakCoupling}(c).

The ED results show several salient features for the evolution of superconducting correlations with interactions. In particular, $s$-wave correlation is largely enhanced when $-0.5 \le V \le 0$ and $U <0$ [Fig.~\ref{fig:correlationsWeakCoupling}(b)]. The $d_{x^2-y^2}$-wave correlation can increase when $V< 0$ and $U > V$ [Fig.~\ref{fig:correlationsWeakCoupling}(d)]. The $p$-wave correlation also can increase when $V<0$ and $U >0$ [Fig.~\ref{fig:correlationsWeakCoupling}(f)]. Other correlations with different pairing symmetries, such as $d_{xy}$ or $s^*$, also can be suppressed or enhanced depending on the interactions [Fig.~\ref{fig:correlationsWeakCoupling}(e) and (g)]. As discussed later, these more exotic superconductivity pairings may be enhanced at larger interaction strengths or at higher doping levels.

In principle, correlation functions computed on a finite-size cluster cannot determine directly the symmetry breaking phases, as we cannot observe true phase transitions but only cross-over phenomena. 
Other approaches such as finite-size extrapolation for all orders are needed to determine the leading instability in the thermodynamic limit\,\cite{ying2014determinant, terletska2017charge, jiang2019superconductivity, PhysRevB.102.041106, jiang2022stripe}, which for ED is not practical beyond 16-site calculations. In our results, however, it is clear that the regions where correlation functions show apparent enhancement or suppression are already reminiscent of the actual phase boundaries obtained by previous fRG studies~\cite{huang2013unconventional}. Our weak-coupling phase diagram at half filling is summarized in Fig.~\ref{fig:correlationsWeakCoupling}(h). Compared to fRG~\cite{huang2013unconventional}, our result includes additionally a PS region, and a region with strong $p$-wave superconducting instability, which is also suggested by mean-field theory~\cite{nayak2018exotic}.

\begin{figure}[!b]
\begin{center}
\includegraphics[width=\columnwidth]{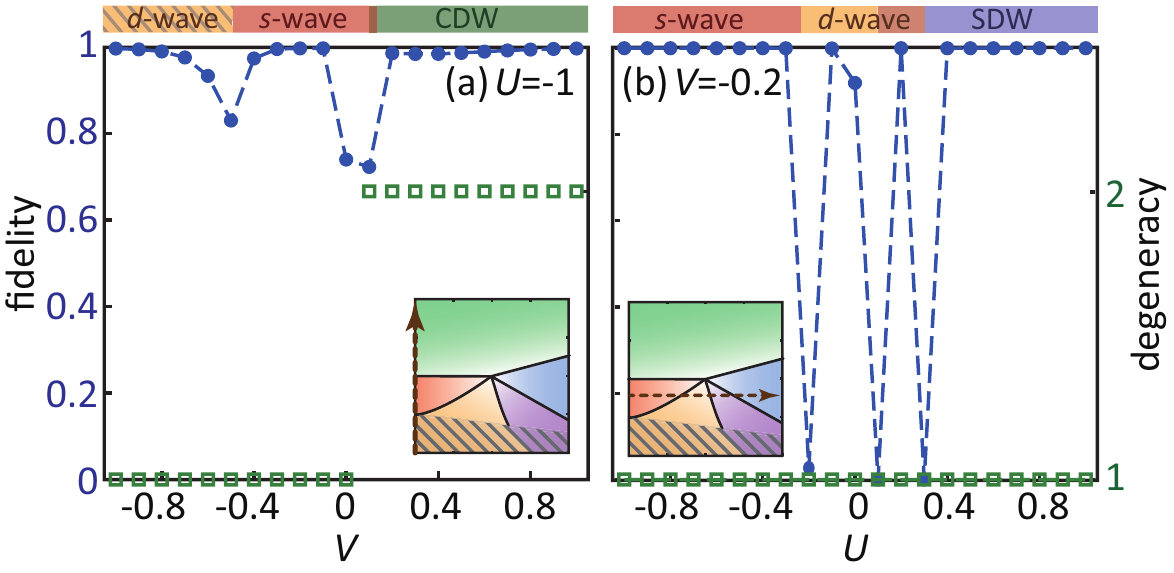}\vspace{-3mm}
\caption{\label{fig:fidelityWeakCoupling}
Quantum fidelity (blue line) and the number of ground state degeneracy (green line) for the extended Hubbard model at half filling. The calculations are performed along different paths shown by arrow lines on the inset weak-coupling phase diagram: (a) $V: -1.0 \rightarrow 1.0 $ at fixed $U=-1.0$, with an interaction step $\Delta V = 0.1$. (b) $U: -1.0 \rightarrow 1.0 $ at fixed $V=-0.2$, with an interaction step $\Delta U = 0.1$. Dips in the fidelity indicate phase transition boundaries.
The calculations are performed on a homogeneous state; gray dashed lines indicate regions where phase separation would occur.
}
\end{center}
\end{figure}

To help determine phase boundaries, we first consider equilibrium calculations of the quantum fidelity $f(\lambda^*,\lambda)$.
Figure \ref{fig:fidelityWeakCoupling}(a) shows the fidelity calculations along the path $V: -1 \rightarrow 1 $ at a fixed $U=-1$, with an interaction step $\delta \lambda = \Delta V = 0.1$. This path is depicted by the arrow line in the figure inset. The fidelity (blue line) exhibits two dips around $V=-0.5$ and 0, and the number of ground state degeneracy (green line) also changes at $V=0$.
We emphasize that our correlation and fidelity calculations are always performed on a ``homogeneous state", instead of a phase-separated one. Our 16-site cluster at a fixed particle sector is too small to show the spatially inhomogeneous PS with both hole-rich and hole-deficient regions. Instead, the PS information is obtained by computing the ground-state energies in different particle sectors (see \hyperref[appendix:PS]{Appendix}). Therefore, although in Fig.~\ref{fig:fidelityWeakCoupling}(a) the fidelity path shows a transition between $d$-wave and $s$-wave superconductivity in the homogeneous state, it is understood that the actual ground state of the system should be phase separated when there is a strong negative $V$. Gray dashed lines are utilized to indicate the PS region in Fig.~\ref{fig:fidelityWeakCoupling}. The same understanding and labeling scheme apply to all the following discussion and figures.

With the above caveat in mind, Fig.~\ref{fig:fidelityWeakCoupling}(a) together with the correlation functions in Fig.~\ref{fig:correlationsWeakCoupling} suggest that the system starts from PS at $(U=-1, V=-1)$, transits to $s$-wave superconductivity around $(U=-1, V=-0.5)$, and enters the $(\pi,\pi)$ CDW state at $(U=-1, V > 0)$.
Since $s$-wave superconductivity and the $(\pi,\pi)$ CDW phase are expected to be degenerate when $V=0$ and $U < 0$, this explains why our calculations appear to show a broader boundary near $V=0$.

Figure \ref{fig:fidelityWeakCoupling}(b) shows the fidelity calculations along the path $U: -1 \rightarrow 1 $ at a fixed $V=-0.2$, with an interaction step $\Delta U = 0.1$. The fidelity exhibits three dips around $U=-0.2$, 0.1, and 0.3, while the ground state remains non-degenerate throughout the whole path (indicated by the arrow line in the figure inset). In accord with our phase diagram in Fig.~\ref{fig:correlationsWeakCoupling}(h), the system starts from the $s$-wave phase at $(U=-1, V=-0.2)$, transits to $d_{x^2-y^2}$-wave superconductivity around $(U=-0.2, V=-0.2)$, and enters the $(\pi,\pi)$ SDW state at $(U=0.3, V = -0.2)$. We note that the fidelity calculation and correlation functions in Fig.~\ref{fig:correlationsWeakCoupling} appear to show additional presence of a $p$-wave state. Since our ground state is a true quantum many-body wavefunction, both leading and sub-leading instabilities may be picked up in the ED calculations. In fact, since $d_{x^2-y^2}$ and $p$-wave superconductivities have similar mean-field energies when $V<0$ and $U >0$\,\cite{nayak2018exotic}, these different orders may coexist in certain regions of the phase diagram.

\begin{figure}[!t]
\begin{center}
\includegraphics[width=\columnwidth]{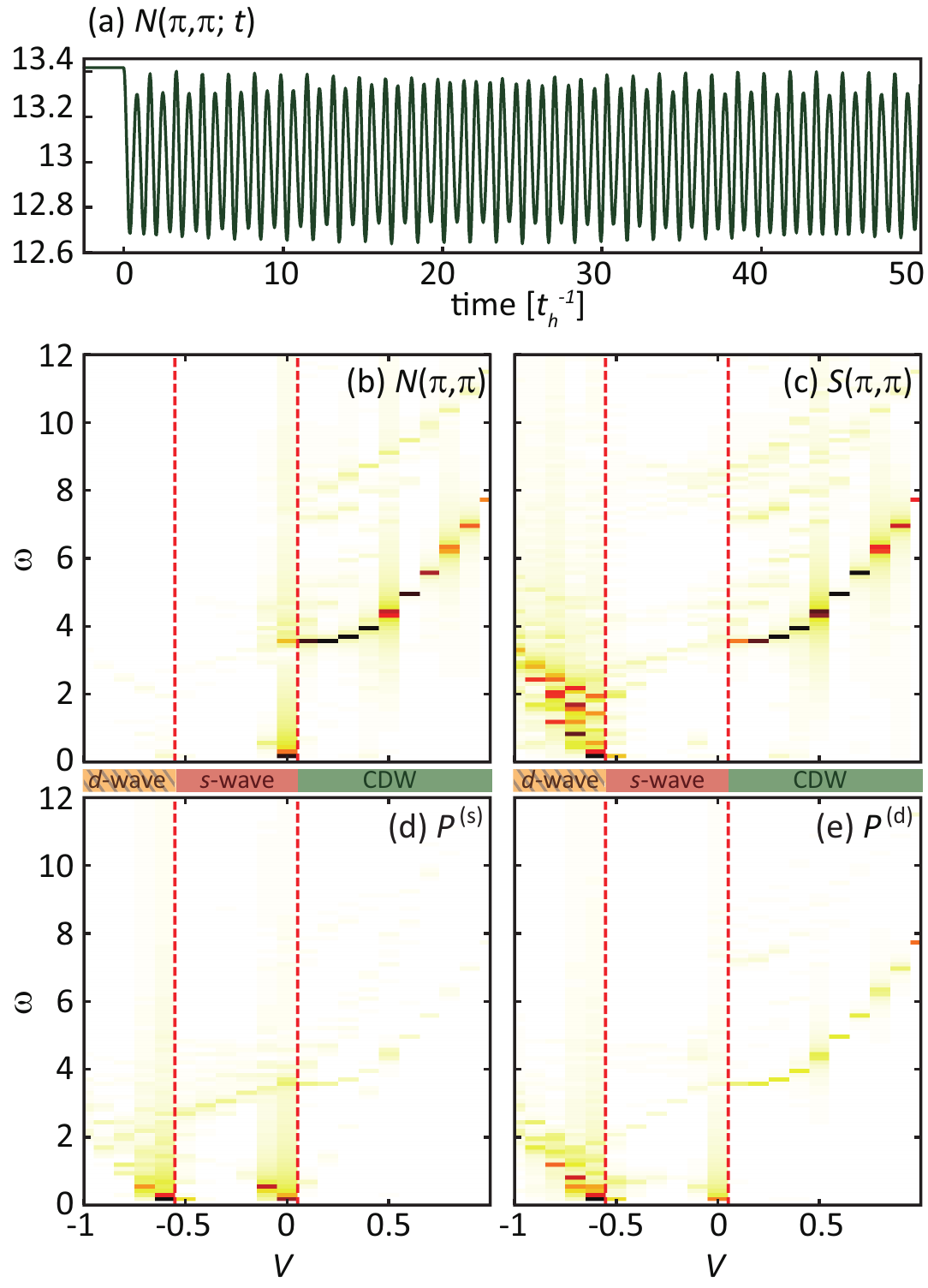}\vspace{-3mm}
\caption{\label{fig:quenchWeakCoupling}
(a) Time evolution of the equal-time correlation function for $(\pi,\pi)$ charge order for the extended Hubbard model at half filling. The initial parameters $(U,V) = (-1, 1)$ at time $t <0$ are quenched to $(U^*,V^*)= (-1, 0.9)$ at $t = 0^+$. (b)-(e) Fourier spectra of the equal-time measurements for $(\pi,\pi)$ CDW, $(\pi,\pi)$ SDW, $s$-wave, $d$-wave superconducting correlations, respectively. The horizontal axis represents $V^*$ after the quench. Vertical dashed red lines indicate phase boundaries near $V=0$ and $V=-0.6$, which agree well with the fidelity calculations.
The calculations are performed on a homogeneous state; gray dashed
lines indicate regions where phase separation would occur.
}
\end{center}
\end{figure}

\begin{figure}[!t]
\begin{center}
\includegraphics[width=\columnwidth]{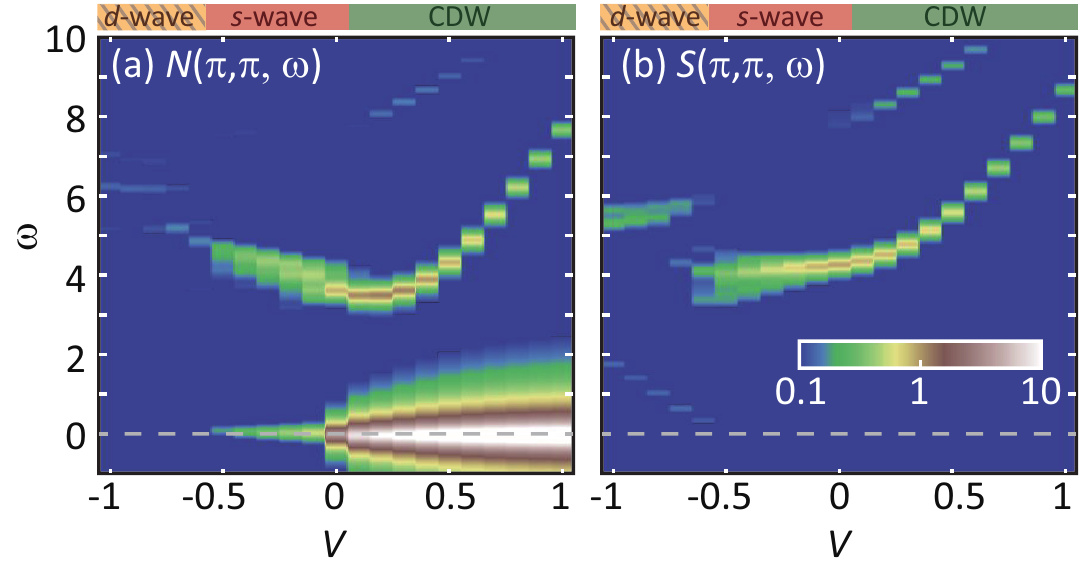}\vspace{-3mm}
\caption{\label{fig:spectrumWeakCoupling}
The charge $N(\mathbf{q},\omega)$ and spin $S(\mathbf{q},\omega)$ dynamical structure factors at $\mathbf{q}=(\pi,\pi)$ for the extended Hubbard model at half filling. The false color intensities are plotted in a log scale. The strong elastic peak in $N(\mathbf{q},\omega)$ for $V \gtrapprox 0$ is caused by doubly degenerate ground states in a $(\pi,\pi)$ CDW phase. The calculations are performed on a homogeneous state; gray dashed lines indicate regions where phase separation would occur.
}
\end{center}
\end{figure}

We next discuss using non-equilibrium quench dynamics to determine equilibrium phase boundaries. Unlike the fidelity metric, which is mainly a theoretical tool, quantum quench of interaction can be prepared for example in optical-lattice or ultrafast experiments\,\cite{meinert2014observation, will2015observation, guardado2018probing, guardado2021quench}, and the subsequent dynamics can be obtained by measuring correlation functions in time domain. Figure \ref{fig:quenchWeakCoupling}(a) shows an example equal-time measurement for charge correlation with $(\pi,\pi)$ ordering $N(\pi,\pi, t)$ [corresponding to the operator in Eq. (3)]. The original equilibrium system at time $t < 0$ is the ground state of Hamiltonian $H_0$ with interactions $(U=-1, V=1)$, which supports a $(\pi,\pi)$ CDW. At time $t = 0^+$, the Hamiltonian is quenched to $H_1$ with interactions $(U=-1, V=0.9)$. $N(\pi,\pi, t)$ then oscillates in time at $t> 0$. For each quenched calculation, we perform Fourier transformations for different equal-time correlations for the operators in Eqs. (3)-(9), in order to extract the characteristic oscillation frequencies associated with quenched dynamics.

Figures \ref{fig:quenchWeakCoupling}(b)-(e) show the resulting Fourier spectra of quenched dynamics for CDW, SDW [for $\mathbf{q}=(\pi,\pi)$], $s$-wave, and $d_{x^2-y^2}$-wave superconductivity [for $\mathbf{q}=(0,0)$], respectively. The horizontal axis represents the post-quench value of $V$. We follow the same path as depicted in Fig.~\ref{fig:fidelityWeakCoupling}(a) inset: $V: -1 \rightarrow 1 $ at a fixed $U=-1$, with a quench step of $\Delta V = 0.1$. The Fourier spectra exhibit distinct behaviors depending on the quenched Hamiltonian. In Fig.~\ref{fig:quenchWeakCoupling}(b), the Fourier spectra of charge correlation $N(\pi,\pi)$ show a clear gap for $V \gtrapprox 0$. The gap size rises with increasing $V$, changing from $\sim 3.7$ at $V = 0.0$ to $\sim 7.7$ at $V = 1$. In Fig.~\ref{fig:quenchWeakCoupling}(c), the Fourier spectra of spin correlation $S(\pi,\pi)$ also show a similar gap for $V \gtrapprox 0$. The spectral intensity of this gap is largely suppressed between $ -0.5\lessapprox V \lessapprox 0$. Noticeably, new low-energy modes emerge below $V \lessapprox -0.5$. In general, the Fourier spectra of quenched $s$-wave [Fig.~\ref{fig:quenchWeakCoupling}(d)] and $d_{x^2-y^2}$-wave [Fig.~\ref{fig:quenchWeakCoupling}(e)] superconducting correlations resemble those of charge and spin correlations.
We note again that the calculations are performed on a homogeneous state; gray dashed lines in the figure indicate regions where phase separation would occur.
Based on the quench behaviors, Figs. \ref{fig:quenchWeakCoupling}(b)-(e) can be separated into three regions [indicated by the vertical dashed red lines]: a phase separated state for $V \lessapprox -0.5$, $s$-wave superconductivity for $ -0.5\lessapprox V \lessapprox 0$, and $(\pi,\pi)$ CDW for $V \gtrapprox 0$.
The quench calculations agree well with the fidelity results, demonstrating the applicability of using a novel non-equilibrium approach to probe equilibrium phase transitions.

To understand the spectral features of quenched measurements, we compute the charge $N(\mathbf{q},\omega)$ and spin $S(\mathbf{q},\omega)$ dynamical structure factors at the ordering vector $\mathbf{q}=(\pi,\pi)$. Here, $|G_{\lambda^*}\rangle$ in Eqs. (10)-(11) is the equilibrium ground state of $H_1$ (the Hamiltonian with the after-quench interactions). As seen in Fig.~\ref{fig:spectrumWeakCoupling}(a), $N(\mathbf{q},\omega)$ shows a clear charge gap for $V \gtrapprox 0$, and the gap behaves similarly as that in Fig.~\ref{fig:quenchWeakCoupling}(b): The gap starts as $\sim 3.7$ at $V=0$ and increases with increasing $V$, and it reaches $\sim 7.7$ at $V=1$. For $V \gtrapprox 0$, the system develops $(\pi,\pi)$ CDW order by spontaneously breaking the discrete charge symmetry, so it lacks a low-energy excitation (Goldstone mode). On the other hand, $N(\mathbf{q},\omega)$ exhibits a strong elastic peak for $V \gtrapprox 0$. This elastic peak in the CDW phase is due to doubly degenerate ground states: one with zero momentum and the other with momentum $(\pi,\pi)$. As seen in Fig.~\ref{fig:spectrumWeakCoupling}(a), $N(\mathbf{q},\omega)$ changes its behavior for $V \lessapprox 0$, where the elastic peak disappears and the charge gap increases with decreasing $V$. The result implies a phase transition near $V=0$. Another phase boundary near $V=-0.5$ can be inferred from $S(\mathbf{q},\omega)$ in Fig.~\ref{fig:spectrumWeakCoupling}(b), where a low energy mode emerges below $V \lessapprox -0.5$. The above results show that the Fourier spectra of equal-time correlations after quantum quench can track the excitation energies in $N(\mathbf{q},\omega)$ and $S(\mathbf{q},\omega)$. However, while dynamical correlations provide information on the underlying broken-symmetry phase (if there is one), they cannot inform directly how close the ground state is to a phase transition, which instead can potentially be inferred by experiments with varying strengths of quenched interaction to examine whether the system resides in proximity to a phase boundary or in a deeply ordered state.

\begin{figure*}[!t]
\begin{center}
\includegraphics[width=16cm]{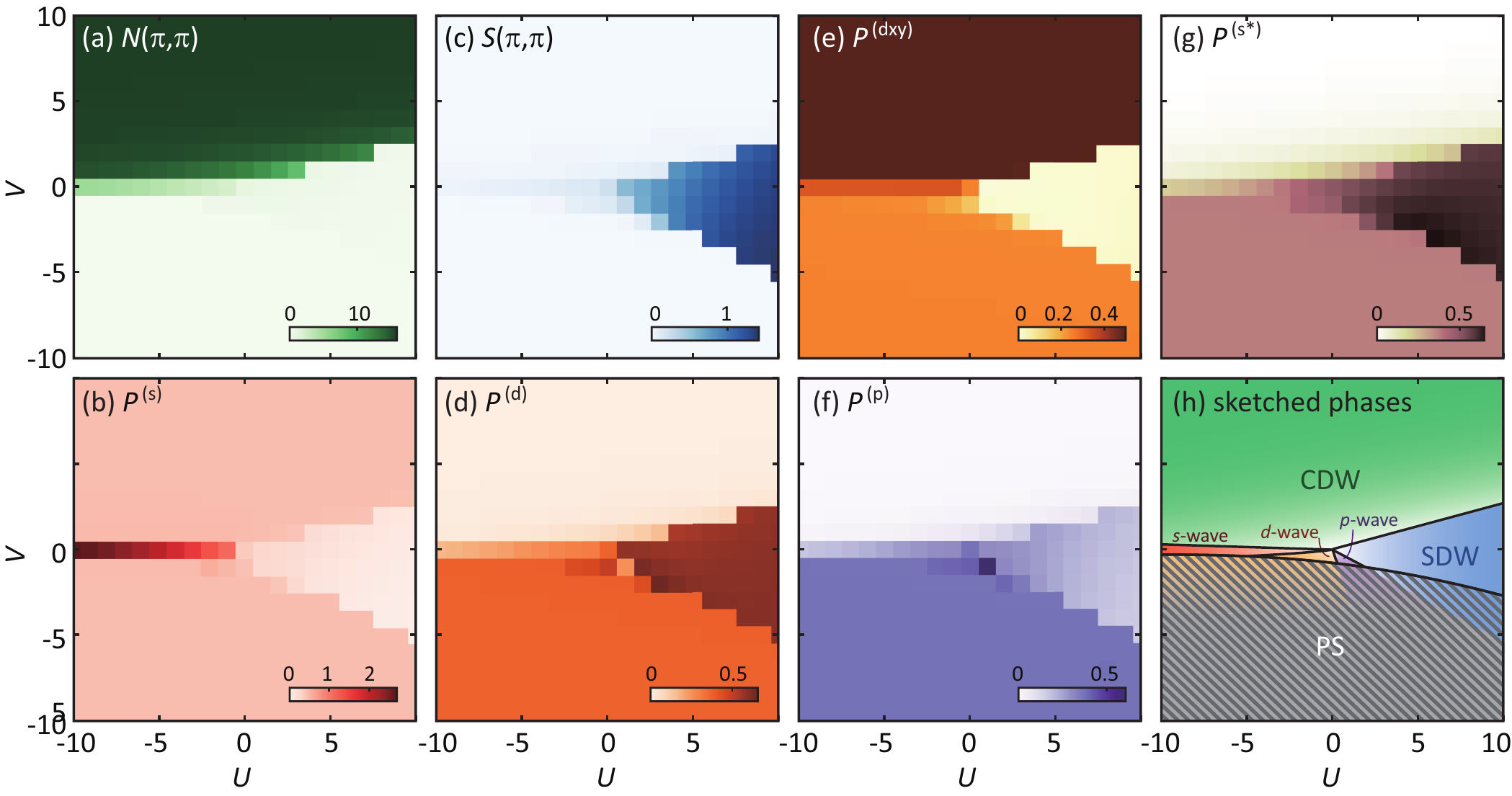}\vspace{-3mm}
\caption{\label{fig:correlationsStrongCoupling} 
Correlation functions for different order parameters computed by exact diagonalization (ED) for half-filled extended Hubbard model in the strong-coupling regime: $|U| \le 10$ and $|V| \le 10$. Panel (h) depicts the corresponding phase diagram with phase boundaries determined by ED calculations of fidelity metric, correlation function, and total energy.
Gray dashed lines indicate regions where phase separation (PS) would occur.
}
\end{center}
\end{figure*}

\begin{figure}[!b]
\begin{center}
\includegraphics[width=0.95\columnwidth]{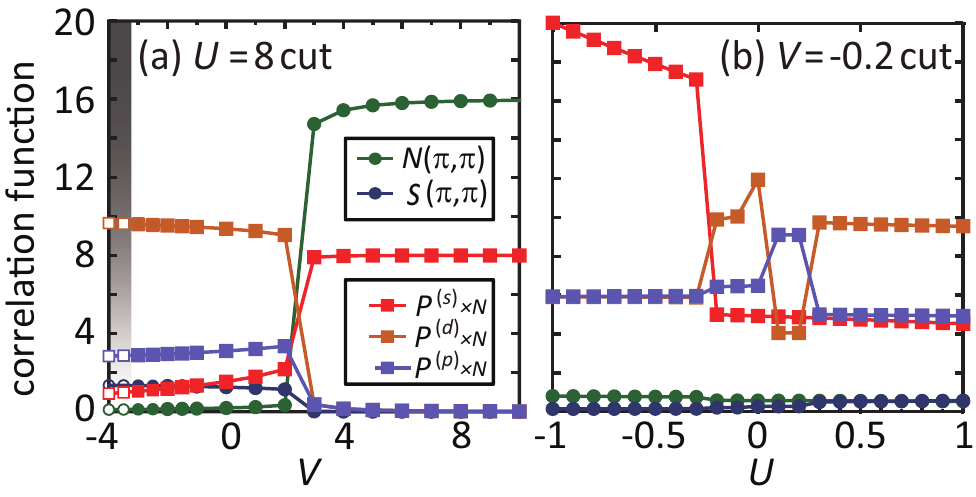}\vspace{-3mm}
\caption{\label{fig:undopedCuts} (a) Evolution of correlation strengths for different order parameters as a function of $V$ for $U=8$ at half filling. The gray area near $V \sim -4$ indicates the phase separation regime (same as the gray dashed regime in Fig.~\ref{fig:correlationsStrongCoupling}). (b) Same as panel (a) but for the evolution along the $U$ direction with $V=-0.2$, which is above the phase separation boundary.}
\end{center}
\end{figure}

\begin{figure}[!b]
\begin{center}
\includegraphics[width=\columnwidth]{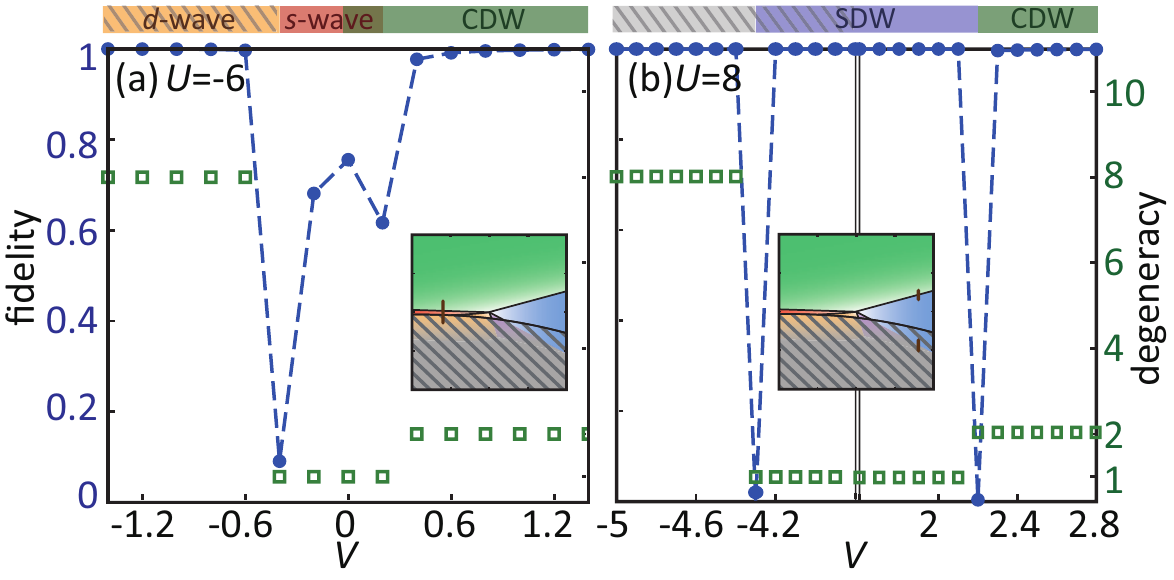}\vspace{-3mm}
\caption{\label{fig:fidelityStrongCoupling}
Quantum fidelity (blue line) and the number of ground state degeneracy (green line) for the extended Hubbard model at half filling. The calculations are performed along different paths shown by arrow lines on the inset {\emph strong-coupling} phase diagram: (a) $V: -1.4 \rightarrow 1.4 $ at fixed $U=6$, with an interaction step $\Delta V = 0.2$. (b) $V: -5 \rightarrow 2.8$ at fixed $U= 8$, with an interaction step $\Delta V = 0.1$. Dips in the fidelity indicate phase transition boundaries.
The calculations are performed on a homogeneous state; gray dashed lines indicate regions where phase separation would occur.
}
\end{center}
\end{figure}

\begin{figure}[!t]
\begin{center}
\includegraphics[width=\columnwidth]{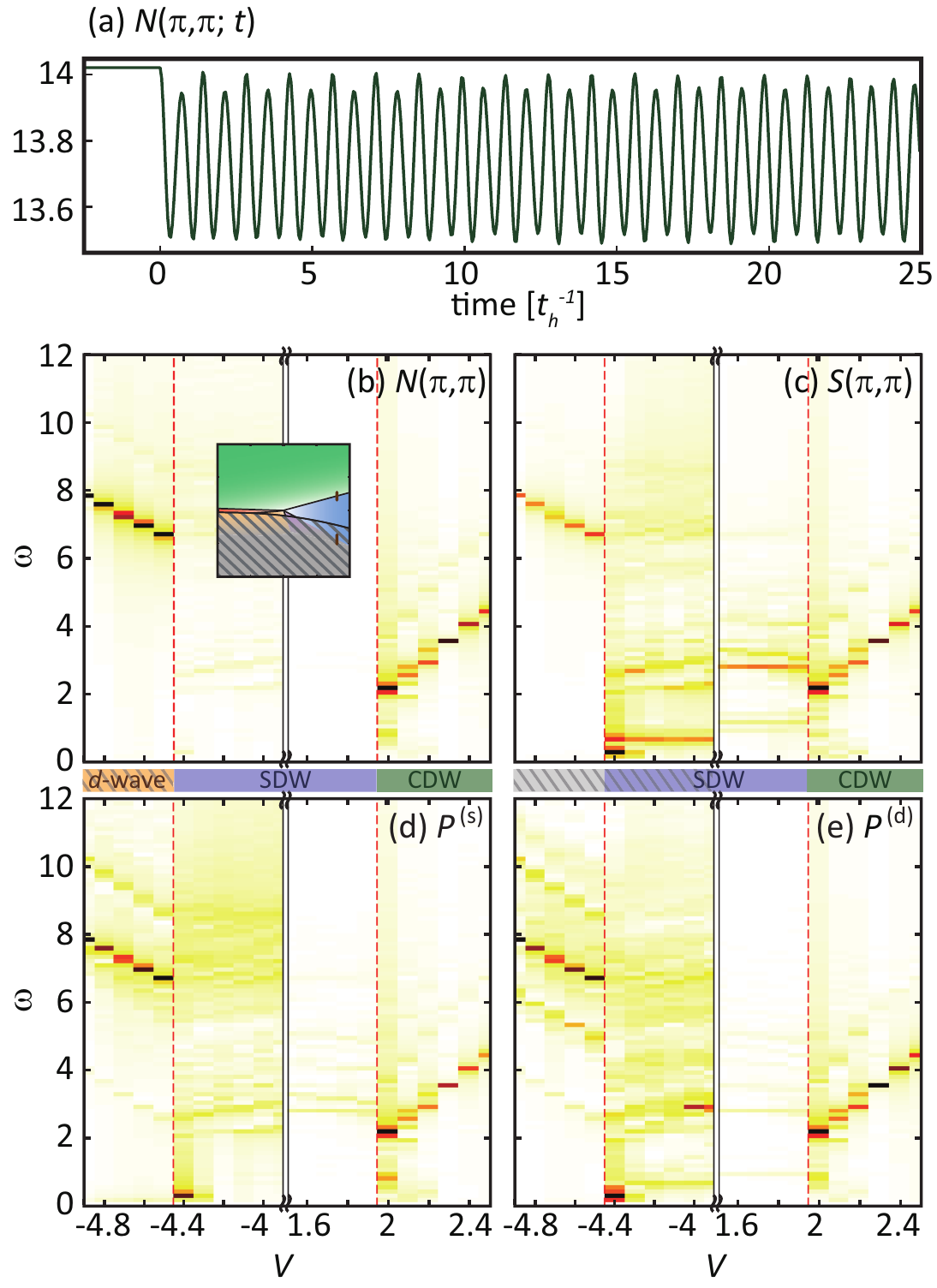}\vspace{-3mm}
\caption{\label{fig:quenchStrongCoupling}
(a) Time evolution of the equal-time correlation function for $(\pi,\pi)$ charge ordering computed for the extended Hubbard model at half filling. The initial parameters $(U,V) = (8, 2.6)$ at time $t <0$ are quenched to $(U^*,V^*)= (8, 2.5)$ at $t = 0^+$. (b)-(e) Fourier spectra of the equal-time measurements for $(\pi,\pi)$ CDW, $(\pi,\pi)$ SDW, $s$-wave, $d$-wave superconducting correlations, respectively. The horizontal axis represents $V^*$ after the quench. Vertical dashed red lines indicate phase boundaries near $V=-4.6$ and $V=2$, which agree with the fidelity calculations.
The calculations are performed on a homogeneous state; gray dashed lines indicate regions where phase separation would occur.
}
\end{center}
\end{figure}

\begin{figure}[!t]
\begin{center}
\includegraphics[width=\columnwidth]{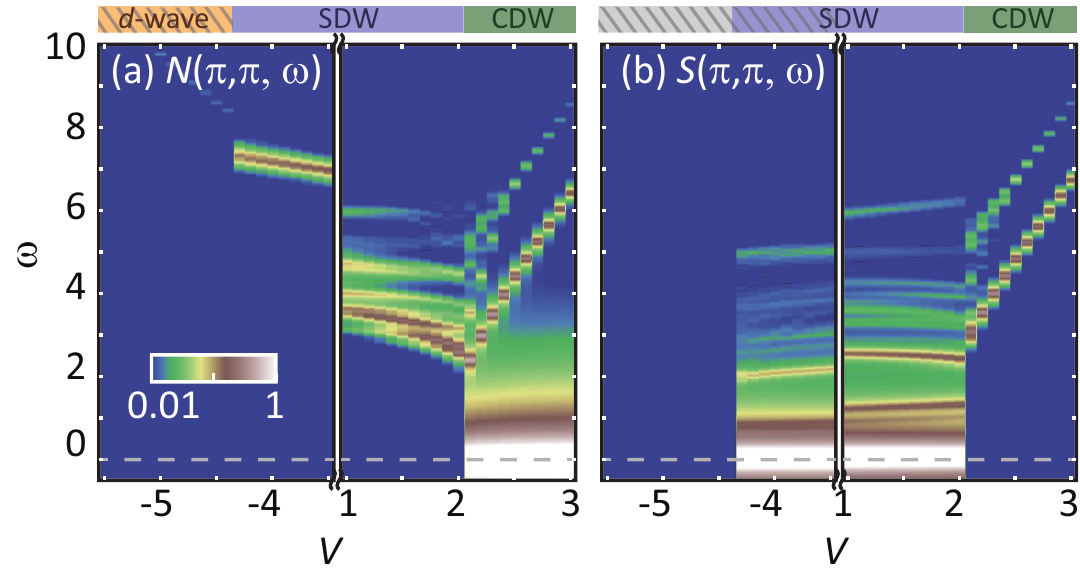}\vspace{-3mm}
\caption{\label{fig:spectrumStrongCoupling}
The charge $N(\mathbf{q},\omega)$ and spin $S(\mathbf{q},\omega)$ dynamical structure factors at $\mathbf{q}=(\pi,\pi)$ for the extended Hubbard model at half filling for strong coupling. The false color intensities are plotted in a log scale. The strong elastic peak in $N(\mathbf{q},\omega)$ for $V \gtrapprox 0$ is caused by doubly degenerate ground states in a $(\pi,\pi)$ CDW phase.
The vertical lines indicate phase boundaries near $V=0$ and $V=-0.6$.
The calculations are performed on a homogeneous state; gray dashed lines indicate regions where phase separation would occur.
}
\end{center}
\end{figure}

\subsection{Strong-Coupling Regime}
After establishing the abilities of using quantum fidelity and quench dynamics to probe phase boundaries in the weak-coupling limit, we now focus on the strong-coupling regime: $|U| \le 10$ and $|V| \le 10$.
Figure \ref{fig:correlationsStrongCoupling} shows equilibrium correlation functions $\langle G_\lambda | \hat O^\dagger \hat O|G_\lambda\rangle$, which exhibit overall four distinct regions: A positive $V > 0$ will favor a $(\pi,\pi)$ CDW [Fig.~\ref{fig:correlationsStrongCoupling}(a)], and together with $U < 0$ it will further enhance the CDW state. In contrast, a predominant positive $U > 0$ will stabilize a $(\pi,\pi)$ SDW [Fig.~\ref{fig:correlationsStrongCoupling}(c)], and the SDW phase boundary along the $V$-axis is enlarged with increasing $|V|$. In the atomic limit, on a half-filled $N$-site square lattice, the energy of the CDW state is $U\times N/2$ (since there are $N/2$ doubly occupied sites), while that of the SDW phase is $V\times 2N$ (since there are $2N$ distinct interaction bonds without double counting). Therefore, the CDW to SDW transition boundary should occur at $U \ge 4V$, which is consistent with our calculations and previous studies\,\cite{dagotto1994superconductivity, onari2004phase,aichhorn2004charge}.

Near $V\sim 0$ and $U < 0$, the system exhibits strong $s$-wave pairing correlations, as indicated by the dark red strip in Fig.~\ref{fig:correlationsStrongCoupling}(b). When $V <0$, the phase diagram has a substantial region dominated by $d_{x^2-y^2}$-wave pairing [Fig.~\ref{fig:correlationsStrongCoupling}(d)].
However, we note that with a predominant negative $V$, the system will tend to be phase separated (into hole-rich and hole-deficient regions)\,\cite{dagotto1994superconductivity}, which again can be understood qualitatively using energy consideration in the atomic limit.
The actual phase boundary between a homogeneous ground state and a phase-separated one can be quantitatively determined from the total energy calculations in different particle sectors (see \hyperref[appendix:PS]{Appendix}). 
Based on the above results, we depict the strong-coupling half-filled EHM phase diagram in Fig.~\ref{fig:correlationsStrongCoupling}(h).

Following the strategy used in the weak-coupling system, we first resort to fidelity calculations to help determine the phases boundaries. For $U < 0$, the ED correlation functions in Fig.~\ref{fig:correlationsStrongCoupling} suggest that the phase diagram consists of three regions, separated by phase boundaries around $V=0$. 
\textcolor{red}{The evolutions of correlation strengths for different order parameters as functions of $V$ (with $U=8$) and of $U$ (with $V=-0.2$) are further displayed in Fig. \ref{fig:undopedCuts}}.
Figure \ref{fig:fidelityStrongCoupling}(a) shows the fidelity (blue line) and the number of degenerate ground states (green line) along the path $V: -1.4 \rightarrow 1.4$ at $U=-6$, as depicted in the figure inset. The fidelity deviates from unity between $-0.6 \lesssim V \lesssim 0.4$, and the number of ground state degeneracy also changes accordingly. Therefore, based on the correlation function and fidelity results, the system begins as PS at $(U =-6, V \lesssim -0.6)$, then transits to the $s$-wave phase at $(U=-6, -0.6 \lesssim V \lesssim 0.4)$, and reaches the CDW state at $(U=-6, V \gtrsim 0.4)$. Compared to the $U=-1$ case in Fig.~\ref{fig:fidelityWeakCoupling}(a), a more attractive $U$ will slightly enlarge the $s$-wave phase boundary along the $V$-axis.

For $U > 0$, the $s$-wave pairing is largely suppressed, and the phase diagram shows three regions consisting of CDW, SDW, and PS. Figure \ref{fig:fidelityStrongCoupling}(b) shows the fidelity and the number of ground state degeneracy along the path $V: -4.8 \rightarrow 2.4$ at $U=8$, as depicted in the figure inset. The fidelity basically remains unity except at the critical values $V=-4.3$ and $2.2$, where the fidelity drops to almost zero, signaling two phase transition boundaries. Therefore, the system begins as PS at $(U =8, V \lesssim -4.3)$, transits to the $(\pi,\pi)$ SDW phase at $(U=8, -4.3 \lesssim V \lesssim 2.2)$, and reaches the CDW state at $(U=8, V \gtrsim 2.2)$. Overall, the phase boundaries suggested by fidelity match the regions where correlation functions change more abruptly.

We next discuss the quench dynamics. Figure \ref{fig:quenchStrongCoupling}(a) shows the equal-time measurement for charge correlation at $(\pi,\pi)$. The original system at time $t < 0$ is the ground state of Hamiltonian $H_0$ with interactions $(U=8, V=2.6)$, which supports a $(\pi,\pi)$ CDW. At time $t = 0^+$, the Hamiltonian is quenched to $H_1$ with interactions $(U=8, V=2.5)$. $N(\pi,\pi, t)$ then oscillates in time at $t> 0$.
Figures \ref{fig:quenchStrongCoupling}(b)-(e) show the resulting Fourier spectra of quenched dynamics for $(\pi,\pi)$ CDW, $(\pi,\pi)$ SDW, $s$-wave, and $d_{x^2-y^2}$-wave superconductivity, respectively. The horizontal axis represents the post-quench value of $V$. Here we follow the same path as depicted in the inset of Fig.~\ref{fig:fidelityStrongCoupling}(b), $V: -4.8 \rightarrow 2.4$ at a fixed $U=8$, with a quench step of $\Delta V = 0.1$. As shown in Fig.~\ref{fig:quenchStrongCoupling}(b), the ``charge gap" behaves quite differently below $V \lesssim -4.6$ and above $V \gtrsim 2.0$. Above $V \gtrsim 2$, the gap is enhanced by $V$, showing an increasingly more robust CDW phase. Between $-4.6 \lesssim V \lesssim 2$, the spectra for the $S(\pi,\pi)$ correlation in Fig.~\ref{fig:quenchStrongCoupling}(c) exhibits low-energy excitation (Goldstone mode) of the spontaneous spin symmetry breaking $(\pi,\pi)$ SDW state.
We note again that the calculations are performed on a homogeneous state; gray dashed lines in the figure indicate regions where phase separation would occur. 
Overall, the quench dynamics spectra can track the charge and spin gaps in the dynamical structure factors $N(\mathbf{q},\omega)$ and $S(\mathbf{q},\omega)$ shown in Fig.~\ref{fig:spectrumStrongCoupling}.

\subsection{Doped Systems}
As shown above, the phase boundaries determined by fidelity and quench dynamics agree well with the correlation function calculations. In hole-doped systems, we thereby use mainly the correlation function results with additional fidelity calculations.
We note that in doped systems, since the momentum dependence and the competition with electron itinerary become more important, it can become challenging to determine the leading instability on a finite-size cluster. Therefore, instead of specifying the exact phases, we will focus on the trends of different order parameters evolving with the interaction strength.
In particular, we concentrate on the heavily overdoped regime with 50\% hole doping. In this case, the system exhibits strong $p$-wave superconductivity correlation in the interaction parameter space $(U, V)= (8, -1)$ relevant to the cuprate superconductors\,\cite{chen2021anomalously}.
In general, the boundaries of phase transition (or cross-over phenomenon on a finite-size cluster) evolve smoothly from the half-filled phase diagram in Fig.~\ref{fig:correlationsStrongCoupling} to the 50\% hole doped system in Fig.~\ref{fig:correlations50}.
The results for other hole fillings are shown in the \hyperref[appendix:doping]{Appendix}.

Figure \ref{fig:correlations50} shows the correlation functions for different order parameters computed at 50\% hole doping in the strong-coupling regime: $|U| \le 10$ and $|V| \le 10$.
When $V > 0$, the correlations are dominated by CDW for $U > 0$ and $s$-wave superconductivity for $U < 0$. $d_{xy}$-pairing correlation is also found to be enhanced when $V > 0$ and $U > 0$, and it may be the leading instability for a small, positive $V$\,\cite{onari2004phase}. When $V < 0$, most of the phase diagram resides in the PS region, especially for $U < 0$. The PS boundary for each doping is determined from the total energy calculations in different particle sectors and is shown in the \hyperref[appendix:PS]{Appendix}. Even if $U$ is positive but with $U < |V|$ (where $V$ is negative), the system remains phase-separated. Interestingly, when $U$ is positive and dominates over $|V|$, both $p$-wave and $d$-wave superconducting correlations can be strongly enhanced. 

The evolution of the correlation strengths for different order parameters as a function of $V$ (at fixed $U = 8$) is shown in Fig.~\ref{fig:correlations50}(h). For the parameter set $U =8$ with a negative $V$ ($\sim -1$) relevant to the cuprates, $p$-wave and $d$-wave correlations have comparable strengths. In general, a repulsive $U$ will favor an inter-site phase-changing superconducting order parameter (like $p$-wave or $d$-wave pairing), and an attractive $V$ can enhance superconducting correlations while suppressing other competing states like CDW\,\cite{peng2023enhanced}. This enhancement of superconductivity by $V$ is beyond the impact of a next-nearest-neighbor hopping \,\cite{jiang2019superconductivity,peng2023enhanced}. An attractive $V$ can further favor $p$-wave superconductivity due to enhanced NN triplet states especially near 50\% hole doping\,\cite{qu2022spin}. In mean-field calculations\,\cite{nayak2018exotic}, the two phases could possibly coexist. This is consistent with the dominance of the $p$-wave instability in 1D systems\,\cite{lin1986condensation,penc1994phase,qu2022spin}, where the $d$-wave competitor is geometrically forbidden. Therefore, it would be interesting to investigate more extensively near this parameter regime both theoretically and experimentally, which may hold the promise of realizing the more exotic $p$-wave superconductivity.

\begin{figure*}[!t]
\begin{center}
\includegraphics[width=16cm]{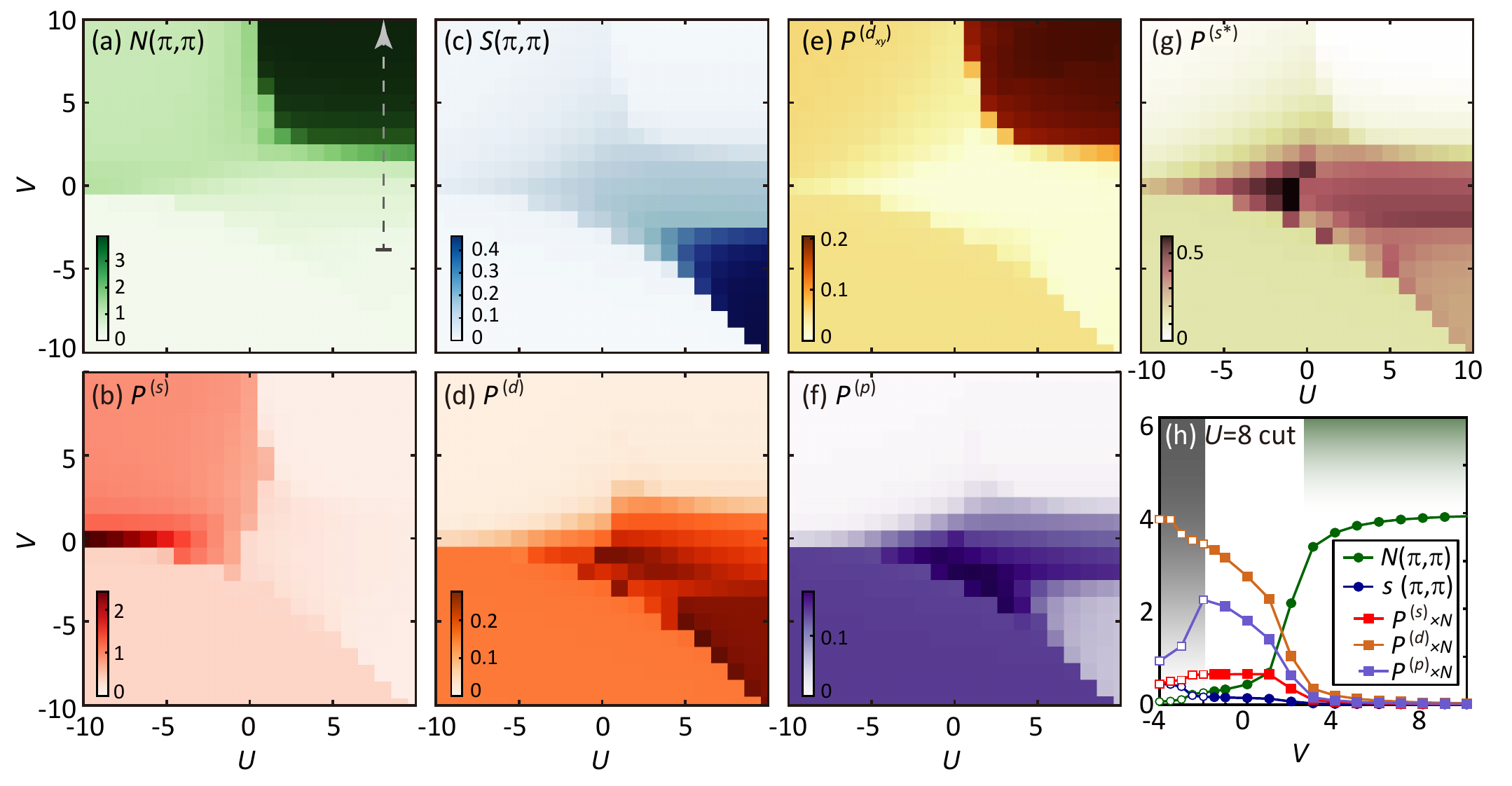}
\caption{\label{fig:correlations50}
Correlation functions for different order parameters computed by exact diagonalization (ED) for 50\% hole-doped extended Hubbard model in the strong-coupling regime: $|U| \le 10$ and $|V| \le 10$. Panel (h) depicts the evolution of correlation strengths for different order parameters as a function of $V$ (at fixed $U=8$). The path is also specified by the dashed arrow in panel (a). Gray shade indicates the phase separation region.
}
\end{center}
\end{figure*}

\section{Conclusion}
We have performed extensive exact diagonalization (ED) calculations to study the extended Hubbard model (EHM) for both attractive and repulsive $U$ and $V$ values, ranging from weak to strong coupling regimes. We have utilized the correlation functions and fidelity metrics to explore how different superconducting phases and their boundaries evolve with the interaction parameters and doping levels. We also have shown that a more novel, experimentally accessible non-equilibrium quench approach can help determine the phase boundary of the equilibrium system. While our ED studies are limited to finite-size clusters, the resulting phase boundaries agree well with other numerical techniques like functional Renormalization Group\,\cite{huang2013unconventional} in the weak-coupling limit. The ED approach remains valid in the strong-coupling limit, and the results should be reliable at a semi-quantitative level. The quenched dynamics in general cannot be easily and accurately computed by other numerical methods, either. Therefore, our exact treatment of the model in finite-size systems also provide systematic benchmark results for other computational techniques\, \cite{linner2023coexistence, jiang2023single} capable of addressing the thermodynamic limit, and an experimental roadmap  for exploring different superconducting order parameters.

Our results indicate that the EHM is a potential platform to realize the more exotic $p$-wave superconductivity in the repulsive-$U$ and (slightly) attractive-$V$ parameter regime, especially when the system is heavily overdoped by hole carriers. This study is timely, since an attractive $V\sim-t_h$ has recently been identified experimentally in doped 1d cuprate chains\,\cite{chen2021anomalously}. The parameter regime $(U, V) = (8t_h, -t_h)$ is also where $p$-wave superconductivity shows strong instability in our calculations. In the actual cuprate materials, the $U$ and $V$ values can behave differently in response to external stimuli, such as strain, pressure, and laser field, so it is likely to induce transitions near a phase boundary to favor certain instability via heterostructure, high pressure, or ultrafast techniques. Studying a possible $p$-wave state in EHM using other computational approaches or experiments can be intriguing and important area for future research.

\section*{ACKNOWLEDGMENTS}
The authors thank Zecheng Shen for the help in preparing the manuscript. W.C.C. and Y.W. acknowledge support from the National Science Foundation (NSF) award DMR-2132338. C.C.C. acknowledges support from NSF Awards OIA-1738698 and DMR-2142801. The calculations were performed on the Frontera computing system at the Texas Advanced Computing Center. Frontera is made possible by NSF Award OAC-1818253.

\bibliography{main}

\clearpage

\section*{APPENDIX}
\appendix

\section{Phase Separation Region}\label{appendix:PS}
\begin{figure*}[!t]
\begin{center}
\includegraphics[width=16cm]{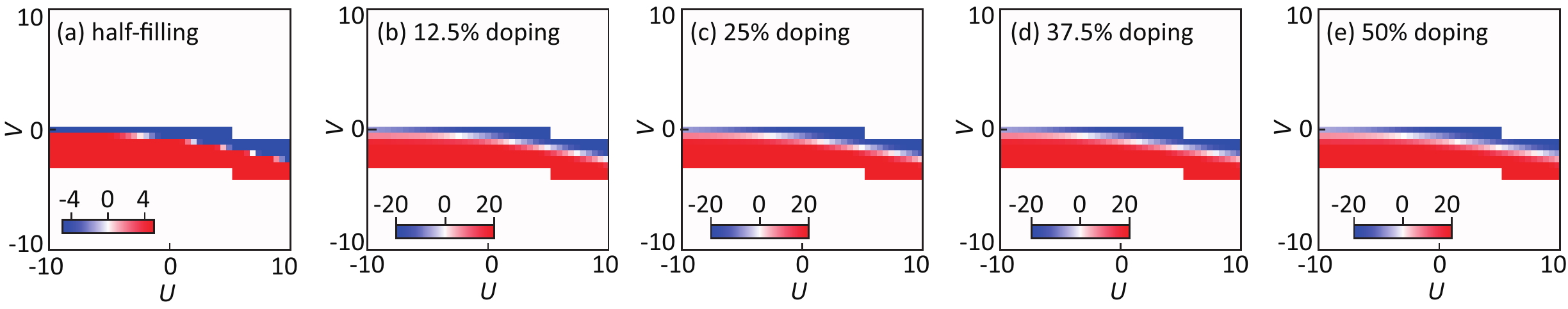}
\caption{\label{fig:energy_PS}
False-color intensity plot of the energy difference $\Delta E$ for determining the phase separation boundary as discussed in the Appendix text. A sign change (from negative to positive) in $\Delta E$ would signal transition to a phase-separated state.
}
\end{center}
\end{figure*}

Our exact diagonalization calculations are performed on a square-lattice $N=16$-site cluster, with a fixed particle sector at a given filling. The cluster size is too small to observe directly the inhomogeneous sparation of hole-rich and hole-deficient regions. Instead, the PS information for a given set of interaction parameters $(U, V)$ can be obtained by computing the total energies in different particle sectors. For example, at the filling with $N_e$ electrons on an $N$-site cluster, the energy difference $\Delta E$ is computed:

\begin{equation}
\Delta E = E(N_e) - \frac{N_e}{2N} \times [E(2N) + E(0)].
\end{equation}
Here, $E(N_e)$ is the energy of the homogeneous ground state with electron occupation number $N_e$; $E(0)$ and $E(2N)$ are respectively the energies of the hole-rich and hole-deficient states on an $N$-site cluster.
Figure \ref{fig:energy_PS} shows the $\Delta E$ false-color intensity plots as functions of $(U, V)$ for different hole-doped systems. A sign change (from negative to positive) in $\Delta E$ would signal transition to a phase-separated state, which helps determine the phase boundary of the PS region.

\section{Correlation Functions at Other Doping Levels}\label{appendix:doping}
\begin{figure*}[!h]
\begin{center}
\includegraphics[width=16cm]{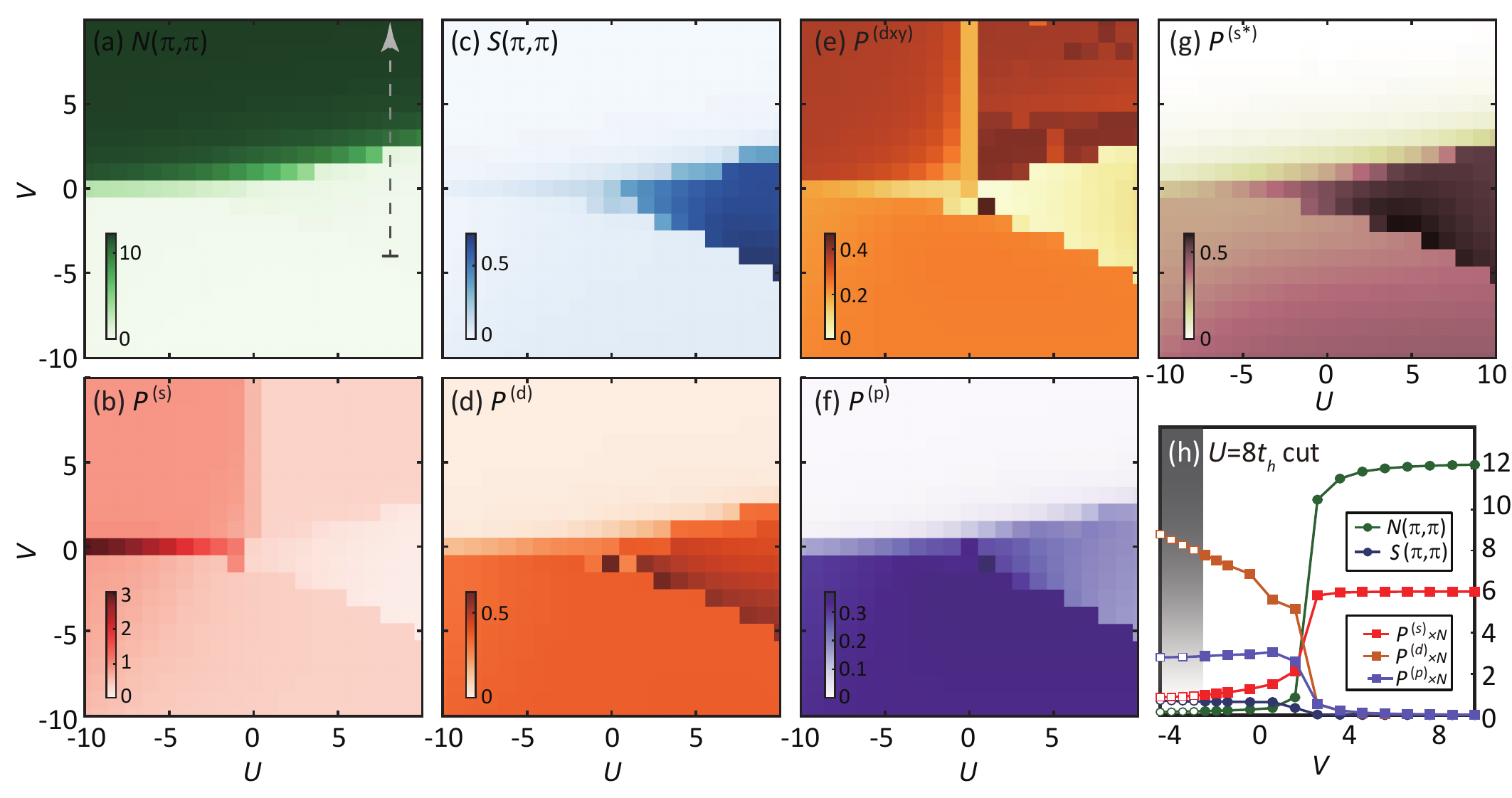}
\caption{\label{fig:correlations12.5}
Correlation functions for different order parameters computed by exact diagonalization (ED) for 12.5\% hole-doped extended Hubbard model in the strong-coupling regime: $|U| \le 10$ and $|V| \le 10$.
Panel (h) depicts the evolution of correlation strengths for different order parameters as a function of $V$ (at fixed $U=8$). The path is also specified by the dashed arrow in panel (a). Gray shade indicates the phase separation region.
}
\end{center}
\end{figure*}

\begin{figure*}[!h]
\begin{center}
\includegraphics[width=16cm]{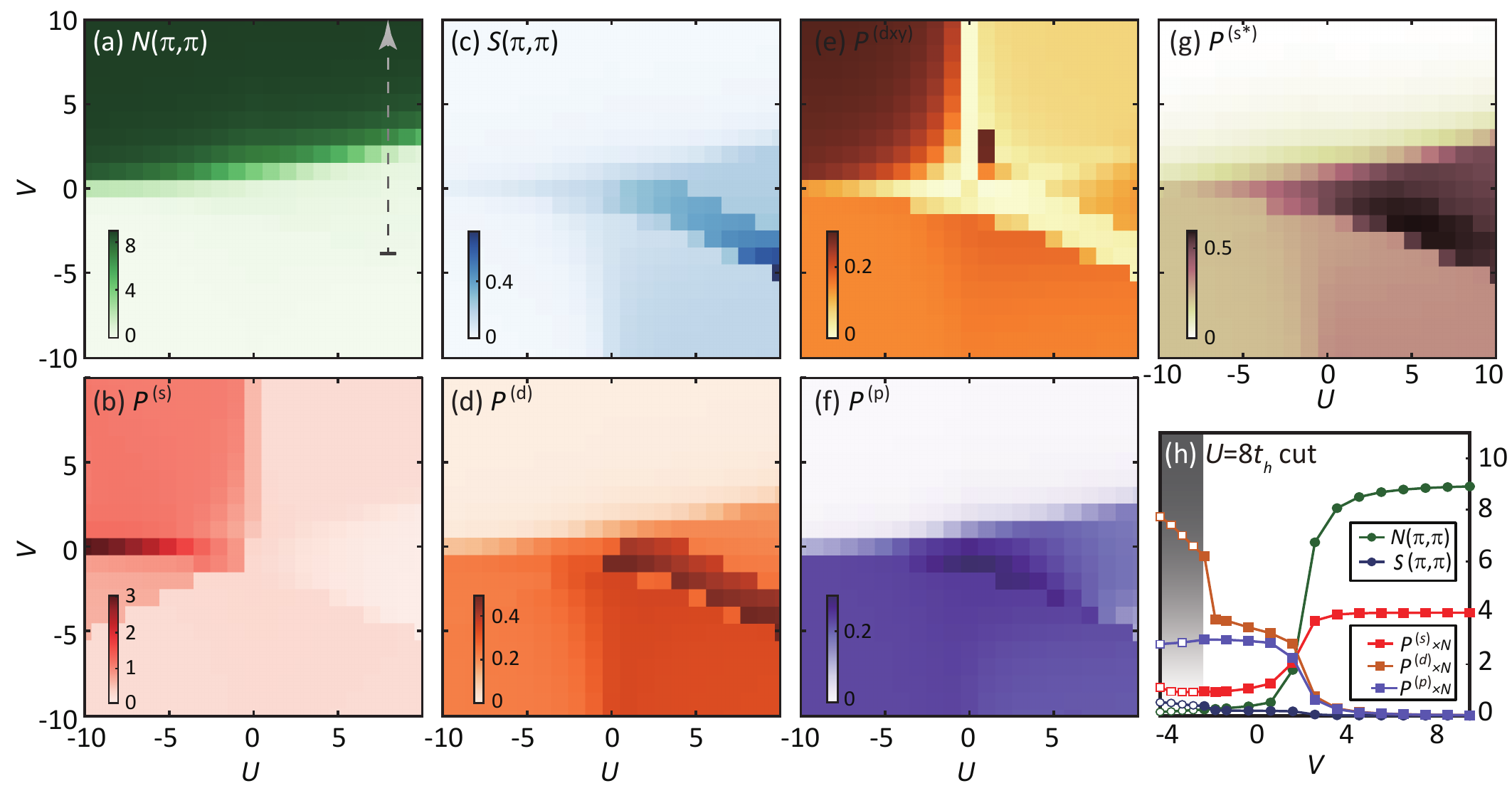}
\caption{\label{fig:correlations25}
Correlation functions for different order parameters computed by exact diagonalization (ED) for 25\% hole-doped extended Hubbard model in the strong-coupling regime: $|U| \le 10$ and $|V| \le 10$.
Panel (h) depicts the evolution of correlation strengths for different order parameters as a function of $V$ (at fixed $U=8$). The path is also specified by the dashed arrow in panel (a). Gray shade indicates the phase separation region.
}
\end{center}
\end{figure*}

\begin{figure*}[!h]
\begin{center}
\includegraphics[width=16cm]{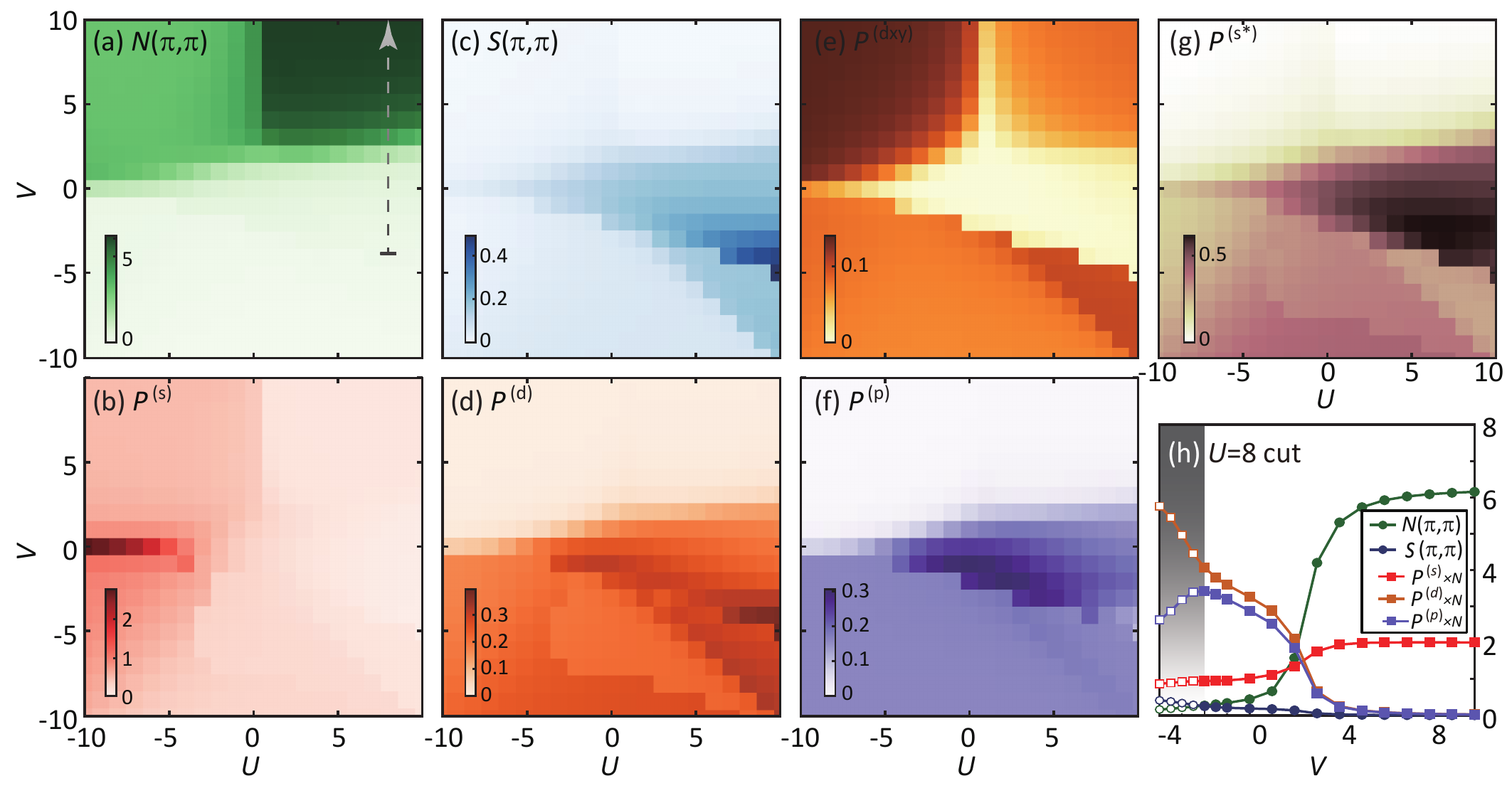}
\caption{\label{fig:correlations37.5}
Correlation functions for different order parameters computed by exact diagonalization (ED) for 37.5\% hole-doped extended Hubbard model in the strong-coupling regime: $|U| \le 10$ and $|V| \le 10$.
Panel (h) depicts the evolution of correlation strengths for different order parameters as a function of $V$ (at fixed $U=8$). The path is also specified by the dashed arrow in panel (a). Gray shade indicates the phase separation region.
}
\end{center}
\end{figure*}

The correlation functions for different order parameters computed in the strong-coupling regime ($|U| \le 10$ and $|V| \le 10$) for 12.5\%, 25\%, and 37.5\% hole-doped systems are shown in Figs. \ref{fig:correlations12.5}, \ref{fig:correlations25}, and \ref{fig:correlations37.5}, respectively.
As discussed above, the PS region for each doping is determined from the energy difference calculation in different particle sectors and is indicated by gray shade in panel (h) of the figure.

\end{document}